\documentclass[jgrga]{agu2001}
\usepackage[dvips]{graphicx}

\authoraddr{Agn\`es Helmstetter, now at Lamont-Doherty Earth Observatory, 
Columbia University (e-mail: agnes@ldeo.columbia.edu)}
\authoraddr{Yan Y. Kagan, Department of Earth and Space Sciences, 
University of California, Los Angeles,
California. (e-mail: ykagan@ucla.edu)}
\authoraddr{David D. Jackson, Department of Earth and Space Sciences, 
University of California, Los Angeles,
California. (e-mail: djackson@ucla.edu)}

\newcommand \be{\begin{equation}}
\newcommand \ba{\begin{eqnarray}}
\newcommand \ee{\end{equation}}
\newcommand \ea{\end{eqnarray}}

\lefthead{Helmstetter, Kagan, and Jackson}
\righthead{Stress transfers and earthquake triggering}

\begin{document}
\setkeys{Gin}{draft=false}
\def\today{\ifcase\month\or
 January\or February\or March\or April\or May\or June\or
 July\or August\or September\or October\or November\or December\fi
 \space\number\day, \number\year}

\hfil PostScript file created: \today

\title{Importance of small earthquakes for stress transfers 
and earthquake triggering}
\author{Agn\`es Helmstetter$^{1,3}$, Yan Y. Kagan$^2$ and David D. Jackson$^{2}$}
\affil{$^1$ Institute of Geophysics and Planetary Physics, 
University of California, Los Angeles, California 90095-1567\\
$^2$ Department of Earth and Space Sciences, University of California,
Los Angeles, California 90095-1567\\
$^3$ Now at Lamont-Doherty Earth Observatory, Columbia University}

\begin{abstract}
We estimate the relative importance of small and large earthquakes for
static stress changes and for earthquake triggering, assuming that 
earthquakes are triggered by static stress changes and that earthquakes 
are located on a fractal network of dimension $D$. 
This model predicts that both the number of events triggered by an
earthquake of magnitude $m$ and the stress change induced by 
this earthquake at the location of other earthquakes increase
with $m$ as $\sim 10^{Dm/2}$. The stronger the spatial clustering, the 
larger the influence of small earthquakes on stress changes at the 
location of a future event as well as earthquake triggering.
If earthquake magnitudes follow the Gutenberg-Richter law with 
$b>D/2$, small earthquakes collectively dominate stress transfer and earthquake
triggering, because their greater frequency overcomes their smaller individual
triggering potential. Using a Southern California catalog, we observe that 
the rate of seismicity triggered by an earthquake of magnitude $m$ increases
with $m$ as $10^{\alpha m}$, where $\alpha=1.05 \pm 0.05$. We also find that 
the magnitude distribution of triggered earthquakes is
independent of the triggering earthquake's magnitude $m$.
When $\alpha \approx b$, small earthquakes are roughly as
important to earthquake triggering as larger ones.
We evaluate the fractal correlation dimension $D$ of hypocenters using two
relocated catalogs for Southern California. The value of $D$ measured
for distances $0.1<r<5$ km is $D=1.54$ for the {\it Shearer et al.}\ [2003] 
catalog and $D=1.73$ for the {\it Hauksson et al.}\ [2003] catalog. The value
of $D$ reflects both the structure of the fault network and the nature of 
earthquake interactions. By considering only those earthquake pairs with 
inter-event times larger than 1000 days, we can largely remove the effects of 
short-term clustering. Then $D\approx2$, close to the value 
$D=2\alpha=2.1$ predicted by assuming that earthquake triggering is due to static 
stress. The value $D \approx 2b$ implies that small earthquakes are as important as larger 
ones for stress transfers between earthquakes, and that considering stress changes 
induced by small earthquakes should improve models of earthquake interactions.
 
\end{abstract}

\begin{article}

\section{Introduction}
Large shallow earthquakes are followed by increased seismic activity 
known as ``aftershocks''. Aftershock sequences of small earthquakes are less 
obvious because the aftershock productivity is weaker, but can be observed 
after stacking many sequences [{\it Helmstetter}, 2003].
Several mechanisms have been proposed to explain earthquake triggering 
due to the static stress change induced by a prior event: 
rate-and-state friction [{\it Dieterich}, 1994], sub-critical crack 
growth [{\it Das and Scholz}, 1981; {\it Shaw}, 1993], viscous relaxation 
[{\it Mikumo and Miyatake}, 1979], static fatigue [{\it Scholz}, 1968], 
pore fluid flow [{\it Nur and Booker}, 1972], or simple sand-pile SOC models 
[{\it Hergarten and Neugebauer}, 2002]. 
{\it Kagan and Knopoff}\ [1987a] proposed that random, Brownian 
motion-like stress fluctuations cause Omori-law aftershock rate decay.
Coulomb stress change calculations have been used to predict the
locations, focal mechanisms and times of future earthquakes (see 
reviews by {\it Harris} [1998], {\it Stein} [1999] and {\it King and Cocco} 
[2001]). Because large earthquakes modify stress 
over a much larger area than smaller ones, and because computing 
Coulomb stress changes requires a good model of slip distribution 
available only for large earthquakes, most studies have neglected the 
influence of ``small'' earthquakes. Most researchers include only the 
mainshock and its largest foreshock or aftershock to predict the location 
of future aftershocks. The success of this approach is significant but 
limited. Only about 60\% of aftershocks are located where the stress 
increased after a mainshock [{\it Parsons}, 2002]; stress shadows (proposed 
decrease of the seismicity rate where Coulomb stress change is negative) are
seldom or never observed [{\it Marsan}, 2003; {\it Felzer et al.}, 2003b]; 
and the correlation of Coulomb stress change with aftershock locations is 
rather sensitive to the assumed slip distribution [{\it Steacy et al.}, 2004]. 
Coulomb stress change from the $M_W$ 7.3 Landers earthquake does not explain 
the triggering of the $M_W$ 7.1 Hector-Mine earthquake 
[{\it Harris and Simpson}, 2002] which occurred 7 years after Landers 
and 20 km away. Several alternative models have been proposed to explain the 
triggering of the Hector-Mine earthquake: viscoelastic effects [{\it Zeng}, 2001], 
dynamic triggering [{\it Kilb}, 2003], or secondary aftershocks 
[{\it Felzer et al.}, 2002]. {\it Felzer et al.}\ [2002] suggested that 
Hector-Mine may have been triggered indirectly by an aftershock of Landers, 
i.e., secondary aftershocks and small earthquakes may be important for stress
triggering. 

A few studies have estimated the scaling of the aftershock number with the 
mainshock magnitude [{\it Solov'ev and Solov'eva}, 1962;
{\it Papazachos et al.}, 1967; {\it Utsu}, 1969; {\it Singh and Suarez}, 1988; 
{\it Yamanaka and Shimazaki}, 1990; {\it Davis and Frohlich}, 1991; {\it Molchan 
and Dmitrieva}, 1992; {\it Shaw}, 1993; {\it Drakatos and Latoussaki}, 2001]. 
In these studies, the ``mainshock'' was defined as the largest event
of a sequence, and ``aftershocks'' were then selected in a space-time
window around the mainshock. These studies proposed that the number of
aftershocks increases exponentially $\sim 10^{\alpha m}$ with the
mainshock magnitude $m$, with an exponent $\alpha$ in the range [0.65, 1]. 
Other researchers [{\it Jones et al.}, 1995; {\it Michael and Jones}, 1998; 
{\it Helmstetter}, 2003; {\it Felzer et al.}; 2004] later suggested that 
the same mechanisms may explain two kinds of triggering: that of aftershocks 
by a previous larger earthquake and that of a large earthquake by a
previous smaller one. Indeed, the distribution of times between an earthquake
(``foreshock'') and a subsequent larger earthquake follows the Omori law as 
for usual aftershocks [{\it Jones and Molnar}, 1979], and the magnitude of triggered 
earthquakes is independent of that for the triggering event [{\it Helmstetter}, 2003]. 
In the work of {\it Helmstetter}\ [2003] and {\it Felzer et al.}\ [2004], 
``aftershocks'' were selected as earthquakes occurring in a space-time window
after a ``mainshock'' (any earthquake not preceded by a larger one), and could 
be larger than the mainshock. Using different methods (see section \ref{seca}), 
{\it Helmstetter}\ [2003] found $\alpha=0.8 \pm 0.1$ and {\it Felzer et al.}\ [2004] 
found $\alpha=1$ for Southern California seismicity. 

Other studies measured the exponent $\alpha$ using a statistical model of 
seismicity [{\it Ogata}, 1989; {\it Kagan}, 1991b; {\it Ogata}, 1992; 
{\it Console et al.}, 2003; {\it Zhuang et al.}, 2004], which assumes that 
any earthquake can trigger other events with a rate which decays in time 
according to Omori's law and which increases with magnitude as 
$\sim 10^{\alpha m}$. {\it Ogata} has coined the term``ETAS'' for ``epidemic 
type aftershock sequence'' to describe this class of models. However, these 
models are now being used to describe the statistics of all earthquakes, not 
just aftershocks. Here we will use the designation ``ETES'' for ``epidemic type 
earthquake sequence''. Optimizing the likelihood of the ETES model gives an 
estimation of the model parameter $\alpha$. When applied to long-time catalogs, 
this method usually gives a smaller value of $\alpha$, in the range 0.5-0.7 
[{\it Console et al.}, 2003; {\it Zhuang et al.}, 2004], than would result by 
simply counting the number of aftershocks. When applied to individual aftershock 
sequences [{\it Guo and Ogata}, 1997], this method gives from one sequence to 
another a large variation of $\alpha$, ranging from 0.2 to 1.9 with a mean value of 0.86.

{\it Yamanaka and Shimazaki}\ [1990] explained the value $\alpha=1$ observed
for interplate earthquakes by assuming that aftershocks
are located on patches of the mainshock fault plane. If the stress 
drop is independent of magnitude, the density of aftershocks on the fault 
plane is a constant. Therefore, the number of aftershocks is
proportional to the area of the mainshock rupture plane, i.e., $\alpha=1$.
{\it Hanks}\ [1992] used the same model to show that, if $\alpha$ is
equal to the exponent $b$ of the Gutenberg-Richter (GR) law, small
earthquakes are just as important as larger ones in redistributing
tectonic forces on faults. {\it Helmstetter}\ [2003] later generalized this model 
by assuming that earthquakes are located on a fractal network of dimension $D$.
This model predicts that stress transfers between earthquakes and earthquake 
triggering scales as $\sim 10^{Dm/2}$. Therefore, if earthquakes are very clustered in
space with $D<2b$, small earthquakes will be more important than larger
ones for transferring static stress between events and thus
for earthquake triggering.

In this work we revisit the work of {\it Helmstetter}\ [2003] by using a 
different method to define mainshocks and aftershocks, and by
discussing the model in more detail. We then test the model by estimating 
the fractal correlation dimension $D$ of earthquake hypocenters, using two 
relocated catalogs for Southern California. We compare the observed value of 
$\alpha$ with the prediction $\alpha=D/2$ [{\it Helmstetter}, 2003].

\section{Scaling of stress transfers and earthquake triggering with
the mainshock magnitude}

We estimate the relative importance of small earthquakes for stress transfers
 and for earthquake triggering. Our model is based on the following assumptions:
\begin{itemize}

\item Earthquakes are triggered by a static stress step induced by the 
mainshock. Triggering 
from dynamic stress changes may have a different scaling with magnitude.
We also assume that the mainshock does not change the stressing rate, i.e., 
we neglect the influence of post-seismic viscous relaxation on the seismicity.
We do not impose a relation between the number of aftershocks and the stress 
change; we simply assume that the rate of aftershocks at point $\vec r$ at 
time $t$ after a stress change $\sigma$ is a function which depends only on 
$t$ and $\sigma$. For instance, in the rate-and-state model [{\it Dieterich}, 1994], 
the instantaneous increase in seismicity rate is proportional to $\exp(\sigma)$, 
but the total number of aftershocks (integrating over time) is proportional 
to the stress change.

\item A triggered earthquake's size is independent of the magnitude 
of the triggering event (``mainshock'') as suggested by [{\it Helmstetter}, 2003].
This implies that the crust is everywhere close to failure, such that 
any small earthquake, triggered by a previous small one,
can grow into an event much larger than its trigger.

\item We consider earthquakes with rupture width smaller than the thickness 
of the seismogenic crust, for which the rupture length $L$ is proportional 
to its width $W$.

\item The rupture length scales with the seismic moment as $L(M) \sim M_0^{1/d}$. 
For earthquakes smaller than the thickness of the 
seismogenic crust and for a constant stress drop, $d=3$.
For larger earthquakes, several models have been proposed, with $d=1$,
$d=2$ or $d=3$ (see {\it Kagan}\ [2004b] for a review). 

\item Earthquakes are located on a fractal network of dimension $D$.
The spatial clustering of seismicity is due both to the geometry of the fault 
network and to earthquake interactions.
Over long time-scales, aftershocks should cover uniformly 
the active fault network, and thus should share the 
spatial distribution of the non-correlated background seismicity.
In our model, we define $D$ as the fractal dimension of the long-term 
time-independent seismicity, not including short-term clustering due
to earthquake interactions.
\end{itemize}

For a constant stress drop, the Coulomb stress change at point $\vec r$ 
due to a finite dislocation of length $L$ and width $W\sim L$ depends 
only on the ratio $r/L$ and the direction $\Phi$ (set of 3 angles) 
between the rupture plane and the fault plane at point $\vec r$ 
on which we compute the stress [{\it Kagan}, 1991c]. This means that 
the average stress change at a distance $r$ proportional to $L$ from 
an earthquake of length $L$ is independent of this earthquake's magnitude.
The only difference between small and large events is that larger 
ones increase the stress over a larger area. We can thus write 
the stress change (tensor) as a function ${\boldmath \sigma}(r/L,\Phi)$.
We first estimate the scaling of stress transfers between quakes
as a function of the magnitude of the event which increases the stress.
The expectation of the norm of the stress tensor induced by an earthquake of 
length $L$, integrating over the location of all earthquakes is given by
\be
\sigma(L) = {\rm E}(|| {\bf \sigma}(r/L,\Phi)||) = \int \limits_{\Phi}
 d\Phi~ \int \limits_0^{\infty} ~|| {\bf \sigma}(r/L,\Phi)|| \, p_r(r)~dr \, ,
\label{sl}
\ee
where $\Phi$ are limits for a set of 3 angles characterizing fault orientation 
(their exact form depend on the parametrization method) and $p_r(r)$ is the 
time-independent density of earthquakes at a distance $r$ from another earthquake. 

For a fractal distribution, 
\be
p_r(r) \sim r^{D-1}~,
\label{pr}
\ee
where $D$ is the correlation dimension of earthquake hypocenters.
The function $p_r(r)$ in (\ref{sl}) describes the long-term spatial
distribution of seismicity, not including clustering due to aftershocks.
Introducing (\ref{pr}) in (\ref{sl}) and using the new variable $x=r/L$,
we have
\be
\sigma(L) = L^D \int \limits_{\Phi} d\Phi~ 
\int \limits_0^{\infty} ~\sigma(x,\Phi)~p_r(x)~dx \,.
\label{sl2}
\ee
The integral in (\ref{sl2}) is now independent of the event's length; the only 
dependence on magnitude is in the factor $L^D$.
The integral is convergent both in the near field (the stress has a 
$r^{-1/2}$ singularity close to the crack tip which is integrable) 
and in the far field if $D<3$ because $\sigma(x)$ decays as $ 1/x^3$ for $x\gg1$. 
Assuming that $L \sim M_0^{1/3} \sim 10^{0.5 m}$ (for a constant 
stress drop and for $L$ proportional to $W$), we can rewrite the average stress 
$\sigma(M_0)$ induced by an earthquake of moment $M_0$ at the location 
of another event (distributed according to $p_r(r)$) as [{\it
    Kagan}, 1994]
\be
\sigma(M_0) \sim M_0^{D/3} \, .
\label{sm}
\ee
If earthquakes are distributed uniformly in a volume ($D=3$), or if we compute
the integral of the stress change induced by a quake on a volume rather 
than at the location of other events, then $\sigma(M_0) \sim M_0$. In other words, 
stress change integrated on a volume is proportional to seismic moment, 
and is therefore dominated by the largest earthquakes in a catalog if 
the exponent $\beta$ of the cumulative moment distribution is
$\beta<1$ (i.e., $b<1.5$ for magnitudes). In contrast, if we compute 
the stress changes at the location of other quakes, and if 
$D \approx 2$, then the increase in the stress change with $M_0$ is 
balanced by the decreased number of events proportional to $M_0^{-2/3}$
with seismic moment (equivalent to GR law with $b=1$). 
In this case $D=3\beta$, small earthquakes are just as important 
as larger ones for stress transfers between earthquakes.

If earthquake triggering is due to static stress changes, the seismicity rate
$N(r,\Phi,t)$ at point $(r,\Phi)$ and time $t$ can be
written as a function $f(\sigma,t)$ of the stress change $\sigma$ and of the 
time $t$ after the stress change. 
The function $f(\sigma,t)$ describes any physical mechanism of earthquake triggering 
due to static stress changes, such as rate-and state friction 
[{\it Dieterich}, 1994], sub-critical crack 
growth [{\it Das and Scholz}, 1981; {\it Shaw}, 1993], viscous relaxation 
[{\it Mikumo and Miyatake}, 1979], static fatigue [{\it Scholz}, 1968], or
pore fluid flow [{\it Nur and Booker}, 1972].
The cumulative number of earthquakes 
$N_{\rm aft}(L)$ triggered by an earthquake of length $L$ in the time window 
[$c$, $T$] is given by the integral over time and space of the seismicity rate
\ba
N_{\rm aft}(L) &=& \int \limits _c ^T dt ~ \int \limits_{\Phi} d\Phi~ 
\int \limits_0^{\infty} ~ f(\sigma,t)~p_r(r)~dr \nonumber \\
&=& L^D \int \limits _c ^T dt ~ \int \limits _{\Phi} d\Phi~ 
\int \limits_0^{\infty} ~ f(\sigma,t)~p_r(x)~dx \, .
\label{nl}
\ea
The minimum time $c$ is introduced to avoid the singularity of the seismicity 
rate at time $t=0$ predicted by some models of stress triggering (e.g., in the 
case of the rate-and-state model [{\it Dieterich}, 1994] when a singularity 
occurs in the stress field). The maximum time $T$ is needed to regularize the 
integral at large times for some models which predict an Omori law decay with 
$p\leq1$ (but we can take $T=\infty$ in {\it Dieterich}'s [1994] model). As for 
the stress change $\sigma(L)$ in (\ref{sl2}), the number of triggered earthquakes 
depends on magnitude only in the factor $L^D$. The scaling of aftershock 
productivity with the size of the triggering earthquake is the same as for 
stress transfers. For seismic moment 
\be
N_{\rm aft}(M_0) \sim M_0^{D/3} \, ,
\label{nM}
\ee
and for magnitude
\be
N_{\rm aft}(m)\sim 10^{\alpha m} \quad \rm{with} \quad \alpha=0.5D \, .
\label{nm}
\ee
The result (\ref{nm}) does not depend on the physical mechanism of earthquake triggering, 
i.e., on the specific form of the function $f(\sigma,t)$ in (\ref{nl}), but is valid 
for any model in which aftershock properties (seismicity rate, location, 
duration of the aftershock sequence) are functions of the static stress change induced by 
the mainshock, and on the time since the stress change, and have no other dependance on the mainshock magnitude.
In the next section, we measure the scaling of the aftershock rate with the 
mainshock magnitude for Southern California seismicity. We then estimate the 
fractal dimension $D$ of the spatial distribution of hypocenters in section 
\ref{secD} and compare the results with the model's prediction (\ref{nm}).

\section{Scaling of earthquake triggering with mainshock magnitude:
observations for California seismicity}

We have measured the average rate of triggered earthquakes
(``aftershocks'') following a ``mainshock'' in Southern California to measure 
the scaling of the number of triggered earthquakes
with the triggering magnitude. We first discuss some potential problems 
in analyzing aftershock properties. We describe the catalog in
section \ref{data}. We then define ``mainshocks''
and ``aftershocks'' and explain our declustering procedure. We present 
our results on the scaling of earthquake triggering with the magnitude 
of the triggering event in section \ref{seca}, and findings on the magnitude 
distribution of triggered earthquakes in section \ref{secpm}

\subsection{Problems with estimating aftershock properties}

Several properties of aftershocks are difficult, if not impossible,
to estimate and may be very sensitive to the parameters of the
declustering procedure. It is particularly difficult to define 
the total number of aftershocks. If earthquake triggering follows 
Omori's law $\sim 1/t^p$, with $p\geq1$ and without cut-off
at short times,  there is an infinite number of aftershocks shortly 
after the mainshock when the seismic network is saturated. 
These aftershocks are not reported in the catalogs. 
For $p\leq1$, there are also infinitely many aftershocks at very long times 
(between an arbitrary time $T$ following the mainshock up to $t=\infty$) which 
are mixed up with non-correlated, background quakes. Therefore, we miss most 
aftershocks if we select earthquakes in seismicity catalogs in a finite time 
window after the mainshock. Because it is impossible to measure the total number 
of triggered events, an alternative solution is to measure the rate of aftershocks 
in a finite time window after the mainshock, when the catalog is complete
and before the rate goes back to the constant background level. If the temporal
variation in the number of triggered events is the same for all mainshock 
magnitudes (as expected if the rate of aftershocks depends only on the stress 
and the time since the stress step), then the scaling of the total number 
of aftershocks (exponent $\alpha$ in (\ref{nm})) can be measured from 
that of the rate of triggered events with the mainshock magnitude.

It is also impossible to distinguish between the ``direct'' (triggered by the
``mainshock'' only) and ``secondary'' aftershocks (triggered by a previous 
aftershock from the mainshock). Because an earthquake is probably not triggered 
by a single prior quake, but rather by the cumulative effect of all previous
events (with a weight depending on the time, distance and magnitude 
of this particular quake), it is impossible to tell which earthquake was triggered 
by which other one. Only a stochastic answer can be obtained, assuming a particular 
model of earthquake interactions: for example, that earthquakes obey the 
Epidemic Type Earthquake Sequences (ETES or ETAS) model [ {\it Zhuang et al.}, 
2004]. 
After estimating the parameters by maximizing the likelihood of the model, 
{\it Kagan and Knopoff}\ [1976] and later {\it Zhuang et al.}\ [2004] 
have proposed a method to calculate the probability that an earthquake 
was triggered by a previous one, or that it was a background event.
However, if the number of secondary aftershocks is proportional to the
number of direct aftershocks, then we can measure the total
(observed) rate of triggered seismicity, including direct and
secondary aftershocks, which will have the same dependence on the
mainshock magnitude as does the number of direct aftershocks [{\it
  Helmstetter and Sornette}, 2003].

Rather than by fitting a multi-parameter stochastic clustering model, such as ETES model, 
we have estimated earthquake clustering properties here directly from 
the earthquake catalog, with some simple selection criteria. 
%In contrast, many
%authors (for example, {\it Kagan and Knopoff}\ [1976] and {\it Zhuang 
%et al.}\ [2004]) assume an ETES type model in order to 
%estimate the $\alpha$ parameter or to select aftershocks. 
Although the ETES type models usually describe earthquake clustering well, 
using them to determine parameters like $\alpha$ can be problematic. 
First, the estimates so determined are not very robust, because the model 
parameters are strongly correlated and poorly resolved. Second, the parameter 
estimates may be biased. Inevitable errors in magnitude, location, 
and temporary incompleteness after larger events [{\it Kagan}, 2004a] 
may each bias earthquake statistics by underestimating clustering at short 
distances and times.
{\it Kagan} [1991b, see p.~142 and his Eq.~36] argues that the value of 
$\alpha$ obtained by maximum likelihood inversion of ETES parameters depends 
on the method used to account for the incompleteness
of short-term aftershocks in a catalog. %%
Furthermore, the ETES models generally assume isotropic clustering, whereas 
earthquakes occur preferentially on spatially clustered faults. 
Neglecting this spatial clustering can lead to underestimating
$\alpha$, because the contribution to the total predicted seismicity rate 
from all small aftershocks better predicts the location of future
aftershocks than does the isotropic contribution from the mainshock. 
A smaller $\alpha$ gives more weight to smaller earthquakes and therefore better 
accounts for the following factors: incompleteness of the catalog after a 
large earthquake, the heterogeneity of the spatial distribution of 
aftershocks, fluctuations of aftershock productivity, and magnitude errors 
(see {\it Helmstetter et al.}\ [2004] for more details).

\subsection{Data \label{data} }

We used the catalog of seismicity for California provided by the
Advanced National Seismic System (ANSS) (available at 
http://quake.geo.berkeley.edu/anss/catalog-search.html). This catalog 
merges data from several networks (for this work, essentially
Southern and Northern California and Nevada). We used earthquakes larger 
than a magnitude threshold $m_d=2$, within the time window 1980/1/1-2004/10/15,
%% updated the catalog to include recent events (parkfield) 
and within the rectangular area $32^{\circ} < \rm{latitude} < 37.5^{\circ}$ and
$-122^{\circ} < \rm{longitude}< -114^{\circ}$. This zone does not 
include Long Valley volcanic seismicity. We checked that the catalog is
complete for $m\ge2$ in this area and time period (except at
short times after a large earthquake). We removed explosions from the 
catalog. 

When available, for large earthquakes, we used moment magnitudes provided by 
the Harvard catalog (available at http://www.seismology.\-harvard.edu/\-CMTsearch.html, 
see {\it Ekstr\"om et al.}\ [2003] and references therein). 
Otherwise we used the magnitudes provided by the ANSS catalog (see 
http://quake.geo.berkeley.edu/\-ftp/pub/doc/\-cat5/cnss.catalog.5 
for more details). The magnitudes for smaller earthquakes are generally 
$m_L$, or local magnitudes estimated from short-period body waves. Our use of 
magnitudes other than moment magnitude is a potential problem, 
because the scaling relationships and simple static stress models we employ 
are based on the seismic moment, which we assume is simply related to magnitude. 
{\it Hutton and Jones}\ [1993] show $m_L$ is a fairly good unbiased surrogate 
for moment magnitude in the magnitude range 4.5 to 6,
but there may be a significant bias for quakes smaller than 4.5. At this point 
we have to assume that any bias does not affect the results much, but in the 
future this assumption needs to be tested using moment magnitudes for smaller 
earthquakes.

\subsection{Selection of mainshocks and aftershocks \label{secm}}

The objective of the declustering procedure is to select as mainshocks
only events not preceded by any large earthquake;
then quakes following mainshocks within a specified time and distance
are identified as aftershocks. 
The space and time windows are chosen in order to minimize the influence
of non-correlated earthquakes. There are always arbitrary choices in a declustering 
procedure, so we tested different values for all adjustable parameters.
As a test we also applied this declustering method to synthetic catalogs
to estimate the accuracy and lack of bias in  results. 

Because we want to investigate the temporal decay of triggered
seismicity and the scaling of aftershock productivity with the 
mainshock magnitude, we want a declustering procedure which makes 
minimal assumptions about the properties of aftershock triggering 
in time, space and magnitude, without introducing a-priori any scaling 
of the number of aftershocks or their duration with 
the mainshock magnitude. We assume that larger earthquakes 
influence seismicity in a wider area, proportional
to the mainshock rupture area [{\it Kagan}, 2002a].

Practically, we select as mainshocks any earthquake of magnitude 
$m_M$ not in the influence zone $R_F\times T_F$
of a previous event of magnitude $m \ge m_M - \Delta m$.
We use generally $\Delta m=1$. This choice ensures that 
aftershocks of the foreshock are negligible compared to
aftershocks of the ``mainshock'' because aftershock productivity
increases rapidly with the mainshock magnitude. We compute distances 
between earthquake epicenters, because depth has a larger error than 
horizontal coordinates. The spatial influence zone $R_F$ of a foreshock 
is defined as
\be
R_F(m)=\max(D_F,~N_{\rm L,F} \times L(m)) \, ,
\label{RF}
\ee 
where the constant $D_F$ accounts for a location's accuracy, $L(m)$ is the 
rupture length [{\it WGCEP}, 2003] of an earthquake of magnitude $m$
\be
L(m) = 0.01 \times 10^{0.5m} \quad \rm{(km)} \, ,
\label{LM}
\ee
and $N_{\rm L,F}$ is an adjustable factor.
Mainshocks are selected with a rectangular area $32.1^{\circ} < 
\rm{latitude} <37.4^{\circ}$ and $-121.9^{\circ} < \rm{longitude}
< -114.1^{\circ}$ slightly smaller (by $0.1^{\circ}$) than the box used 
to select foreshocks and aftershocks so as to avoid finite size effects.

We then select as aftershocks of a mainshock all quakes of magnitude 
$m\geq m_d$ in the influence zone $R_A\times T_A$ of the mainshock,
or of a previous aftershock of the mainshock, even if they are larger
than the mainshock. The size of a cluster can thus increase with time,
due to aftershock diffusion or due to secondary aftershocks, as in
{\it Reasenberg}\ [1985]. We use a space window
\be
R_A(m)=\max(D_A,~N_{\rm L,A} \times L(m)) \, .
\label{RA}
\ee
We use parameters $D_A$ and $N_{\rm L,A}$ in (\ref{RA}) generally smaller
than the parameters $D_F$ and $N_{\rm L,F}$ used to define mainshocks,
because we do not want to consider as a mainshock any earthquake
influenced by a previous one, and we do not want to include as aftershocks 
distant quakes mixed with the background.
Thus, our declustering method has 7 adjustable parameters: $D_A$,
$D_F$, $N_{L,A}$, $N_{L,F}$, $T_A$, $T_F$ and $\Delta m$. 
With our algorithm, a small fraction of earthquakes are considered as 
aftershocks of several mainshocks.
For instance, Landers aftershocks are also considered as aftershocks of 
a $m=2.8$ earthquake which occurred 6 hours before Landers. 

Our method is very similar to the one of {\it Reasenberg}\ [1985], except 
that we do not require that aftershocks be smaller than the mainshock, 
and we do not include a priori hypotheses on either the scaling of aftershock 
productivity with the mainshock or on the temporal decrease of
the seismicity rate with time. By contrast, {\it Reasenberg}\ [1985] assumes that 
aftershock productivity increases as $10^{2m/3}$ 
and that the rate of aftershocks follows the Omori law with $p=1$.

The main differences between this work and the previous analysis of
[{\it Helmstetter}, 2003] are that {\it Helmstetter}\ [2003] used
a constant value for $R_F=50$ km (independent of the foreshock magnitude), 
{\it Helmstetter}\ [2003] also used $\Delta m=0$, and selected as aftershocks 
earthquakes in the influence zone of the mainshock only
(not in the influence zone of a previous aftershock). We also use here a 
different catalog from [{\it Helmstetter}, 2003] (with
different minimum magnitude, space and time windows).

Our definition of triggered events is also very similar to that of 
{\it Felzer et al.}\ [2004], but the parameters (minimum magnitude, 
time interval) and the method used to compute $\alpha$ differ.
The only difference in the algorithm of aftershock selection is that 
{\it Felzer et al.}\ [2004] selected aftershocks within the influence zone 
of either the mainshock or of its largest aftershock, while we 
select aftershocks in the influence zone of both the mainshock and all its 
aftershocks. This difference should have only a minor effect, 
as aftershock diffusion is very weak if there is any at all [{\it Helmstetter 
et al.}, 2003]. Secondly, if a mainshock triggers a larger event, most of 
the following events will occur in the influence zone of the largest earthquake, 
which generally includes the influence zone of the previous ``mainshock''.

\subsection{Scaling of earthquake triggering with the magnitude of the
triggering events \label{seca}}

\begin{figure}
\begin{center}
\includegraphics[width=1\textwidth]{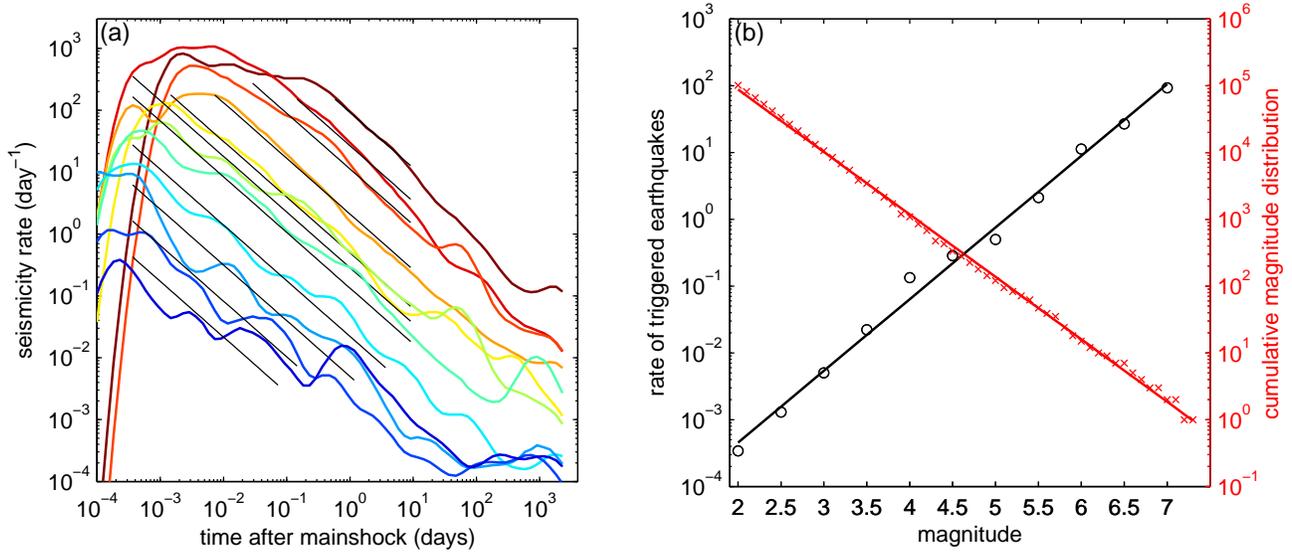}
\caption{\label{figa1}
(a) Average rate $\lambda(m_M,t)$ of $m\geq3$ earthquakes
as a function of the time $t$ after the triggering earthquake, for different 
values of the magnitude $m_M$ of the triggering earthquake
increasing from 2 to 7 with a step of 0.5 from bottom to top. The roll-over 
at short times is due to the incompleteness of the catalog following a 
strong mainshock.
Black lines show the fit of $\lambda(m_M,t)$ by $K(m_M)/t^{0.9}$
in the time interval [$t_{\rm min}(m_M)$, $t_{\rm max}(m_M)$],
where $t_{\rm min}(m_M)=0.0003$ day (26 sec) for $m_M\leq4.5$ mainshocks. 
For larger mainshocks $t_{\rm min}(m_M)$ is estimated using expression (\ref{mc}) 
as the time when the catalog is complete for $m\geq3$.
The maximum time is either 10 days or the time when the seismicity rate decays 
below $\lambda_{\rm min}=0.01$ day$^{-1}$.
(b) Aftershock productivity $K(m_M)$ as a function of $m_M$ 
(circles) and cumulative magnitude distribution $P(m)$ (crosses).
Solid lines are linear regressions of $K(m_M)$ and $P(m)$ with exponent 
respectively equal to $\alpha=1.07$ and $b=0.94$.}

\end{center}
\end{figure}

We stack all aftershock sequences which have the same mainshock
magnitude $m_M$, for each class of mainshock magnitude ranging
from 2 to 7 with a bin size of 0.5. 
We use a kernel method [{\it Izenman}, 1991] to estimate the seismicity rate, by 
convolving  the logarithm of earthquake times 
with a gaussian kernel of width $h=0.1$ for $m_d=2$ and $h=0.2$ for $m_d=3$.
The results for $D_F=3$ km, $D_A= 2$ km,
$N_{L,A}=2$, $N_{L,F}=3$, $T_F=1$ yr, $\Delta m=1$ and $m_d=3$ 
are shown in Figure \ref{figa1}. 
In choosing this parameter, we have checked that our ``mainshocks'' are
almost uniformly distributed in time, and are therefore not strongly 
influenced by other earthquakes. 
%We have, however, a problem after Landers,
%because a few Landers aftershocks occurred up to 1000 km from the epicenter,
%outside our defined influence zone of $R_F=179$ km. Some of these distant 
%aftershocks are thus considered ``mainshocks'' by our declustering algorithm. 
%Therefore, the number of aftershocks from these earthquakes can be over-estimated 
%because aftershocks are mixed up with direct aftershocks of Landers. We 
%have checked that this effect does not significantly affect the results by looking 
%at the time interval 1980-1992.45; it gives almost the same results (see Table 1).

For each mainshock magnitude, the aftershock rate decays with the time since 
the triggering event approximately as 
\be
\lambda(t,m_M)={K(m_M) \over t^p} \, ,
\label{lambda1}
\ee
with $p\approx0.9$. This expression (\ref{lambda1}) is only correct 
in a finite time window [$t_{\rm min}(m_M)$, $t_{\rm max}(m_M)$], 
for which the catalog is complete above $m_d$ and the proportion of 
non-correlated events is negligible.
The Omori law has been reported in many observations of aftershock sequences
(see {\it Utsu et al.} [1995] for a review), and can be derived from several
physical models of earthquake triggering. 
For example, Omori decay (\ref{lambda1}) with an exponent 
$p\approx 0.9$ can be reproduced by the rate-and-state model
of {\it Dieterich} [1994] using a non-uniform stress distribution. 
Due to our rule of aftershocks and mainshocks selection, the seismicity rate
after the mainshock is much larger than before. We conclude that, when Omori 
law decay with $p\approx 0.9$ is observed, the earthquakes which 
we define as ``aftershocks'' are probably causally related to (triggered by) 
the ``mainshock'' rather than simply correlated to it.
Based on analyzing the magnitude distribution for several aftershock sequences 
in Southern California,
{\it Helmstetter et al.}\ [2004] have proposed a relation between the magnitude 
of completeness $m_c(t,m_M)$ as a function of the mainshock magnitude and of
the time (in days) since the mainshock
\ba
m_c(t,m_M) = m_M - 4.5 - 0.76 \log_{10}(t) \nonumber \\
{\rm{and}} \quad m_c(t,m_M) \geq 2.
\label{mc}
\ea 
Earthquakes smaller than this threshold are generally not detected due 
to overlapping of seismic records and saturation of the network.
Of course some fluctuations of $m_c$ occur from one sequence to another, 
but this relation gives a good fit to all 
sequences we analyzed within $0.3$ magnitude units. For Landers $m=7.3$, 
expression (\ref{mc}) predicts that the completeness magnitude recovers 
its usual value $m_d=2$ about 10 days after the mainshock.

For each value of $m_M$, we fit the rate of aftershocks by (\ref{lambda1}) 
with $p=0.9$, in the time interval [$t_{\rm min}(m_M)$, $t_{\rm max}(m_M)$],
to estimate $K(m_M)$.
The minimum time $t_{\rm min}(m_M,m_d)$ is the time after which the catalog is
complete for $m\geq m_d$, estimated using (\ref{mc}). Parameter 
$t_{\rm max}(m_M)$ is either fixed to 10 days or given by the condition 
$\lambda(t,m_M) >\lambda_{\rm min}$, where $\lambda_{\rm min} =0.01$ day$^{-1}$ is the
seismicity rate below which that rate does not obey Omori's law
(\ref{lambda1}). At large times, the seismicity rate goes to a constant level, 
due to the background rate or the influence of non-correlated aftershock 
sequences (see Figure \ref{figa1}). For small $m_M$, the seismicity rate 
increases at large times because the cluster size increases with time, 
as we enlarge the cluster with new earthquakes related to previous ones.
The peak at $t\approx 7$ years for $6\leq m_M<6.5$ and $7\leq m_M<7.5$ is due 
to the $M=7.1$ 1999 Hector-Mine aftershock sequence, which is included in the 
clusters of the 1992 Landers $M=7.3$ and Joshua-Tree $M=6.1$ earthquakes.

Figure \ref{figa1}b shows that aftershock productivity increases exponentially 
with mainshock magnitude.
The value of $K(m_M)$ (representing the seismicity rate per day for $m \geq m_d$ at 
time $t=1$ day after a mainshock of magnitude $m_M$) is given by 
\be
K(m_M)=K_0~10^{\alpha m_M}~10^{-bm_d}~,
\label{Kmm}
\ee
with $K_0=0.0033$ day$^{p-1}$ and $\alpha =1.07$. 
We recover almost the same $\alpha$ value as {\it Felzer et al.} [2004], who
uses a similar method of aftershock selection with parameters
$\Delta m=0$, $T_F=30$ days, $T_A=2$ days, $N_F=N_A=2$, $D_F=D_A=5$ km 
and two values of $m_d=3.5$ and $m_d=4.5$. They measured $\alpha$
by estimating the scaling of the total number of aftershocks in a
2-day period with the mainshock magnitude.
This value of $\alpha$ is close to the GR $b$-value
(see Figure \ref{figa1}b). The $b$-value measured by mean-square linear regression 
of the cumulative distribution for $m\geq2$ in the time window 1980-2004 gives
$b=0.94$. A maximum likelihood method gives $b=1.03$. This value of $b \approx \alpha$ 
implies that all earthquakes in a given magnitude range collectively attain the 
same importance for earthquake triggering, as suggested previously by {\it Agnew and Jones}\ 
[1991], {\it Michael and Jones}\ [1998] and {\it Felzer et al.} [2002, 2004]. 
For $\alpha=b$, the increase of aftershock productivity 
with $m_M$ compensates for the decreased number of earthquakes with magnitude.
Even if an earthquake of magnitude $m=7$ triggers on average $10^5$ times more aftershocks
than a magnitude 2, an earthquake of any magnitude is as likely to be triggered by 
a $m=7$ earthquake as by a $m=2$, because there are $10^5$ more $m=2$ than 
$m=7$ earthquakes in the catalog. 

\begin{figure}
\begin{center}
\includegraphics[width=1\textwidth]{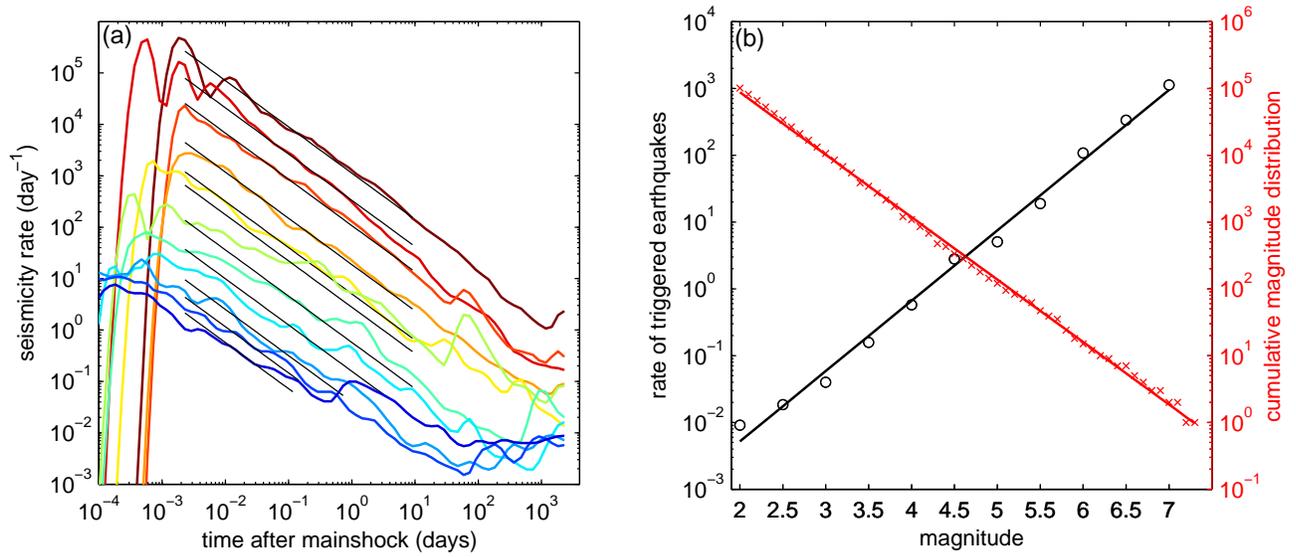}
\caption{\label{figa2}
Same as Figure \ref{figa1} except that we have used $m_d=2$ and we have corrected 
the seismicity rate for missing early aftershocks using (\ref{mc}) (assuming GR law with $b=1$).
We fit the seismicity rate in the time interval $0.002<t<10$ days and 
for $\lambda(t,m_M) > 0.1$ day$^{-1}$. 
The fit of $K(m_M)$ give $K_0=0.0041$ day$^{p-1}$ and $\alpha=1.05$.} 
\end{center}
\end{figure}

The roll-off of the seismicity rate at times $t<t_{\rm min}(m_M)$ can
be explained by catalog incompleteness at shorter times [{\it Kagan}, 2004a].
To check this, we have corrected the seismicity rate using GR law with 
$b=1$ to estimate the rate of seismicity for $m \geq m_d$ from the
observed rate of events for $m \geq m_c(m_M,t)$. We have removed from the
catalog all events with $m< m_c(m_M,t)$, measured the rate of 
seismicity $\lambda_{m\geq m_c}(t,m_M)$ as a function of time and mainshock 
magnitude, and then estimated the rate of $m \geq m_d$ earthquakes by
\be
\lambda_{m\geq m_d}(t,m_M)=\lambda_{m\geq m_c}(t,m_M)~10^{m_c(t,m_M)-m_d}~.
\label{lmc}
\ee
The results are shown in Fig. \ref{figa2}: the rate of seismicity now follows 
the Omori law for $t \geq 2\times 10^{-3}$ days for all mainshock magnitudes.
The cut-off at $t \approx 10^{-3}$ days (86 sec) is probably due to the fact 
that earthquakes at such small times cannot be distinguished from the mainshock 
and are therefore not reported in the catalog.

We have tested different values of the parameters of aftershock and mainshock 
selection (see Table 1). The fitting interval ($t_{\rm max}$ and $\lambda_{\rm min}$)
was adjusted so that the seismicity rate for $t<t_{\rm max}$ 
and $\lambda(t)>\lambda_{\rm min}$ decays according to Omori's law with $p\approx0.9$,
as expected for triggered earthquakes.
All tests, for reasonable values of $D_A$ smaller 
than the location accuracy (2 km for A-quality locations), give $\alpha=1.05 \pm 0.05$.
There are however large variations of $K_0$ as a function of the parameters 
of aftershock and mainshock selection. 
The value of $\alpha$ exceeds 
the value $\alpha=0.8\pm0.1$ found by {\it Helmstetter}\ [2003].
This discrepancy is due to differences in the selection of aftershocks and 
mainshocks. {\it Helmstetter}\ [2003] used a constant value of $R_F=50$ km, 
independent of the ``foreshock'' magnitude. This value   $R_F=50$ 
is too small to exclude distant Landers aftershocks. {\it Helmstetter}\ [2003] 
also used a larger value of $t_{\rm max}=1$ yr, independent of $m_M$, so that 
a significant fraction of background events may have contaminated aftershocks 
of small mainshocks.

We have applied the same method (selection of mainshocks and aftershocks,
correction for undetected early aftershocks, and fit by Omori's law)
to synthetic catalogs generated with the ETES model. In this model, 
the seismicity rate $\lambda(t,\vec r,m)$ is the sum of a constant 
background term $\mu$ and of aftershocks from all previous earthquakes:
\ba
\lambda(t,\vec r,m) &=& p_m(m) \Big[\mu + 
\sum_{t_i<t} k10^{\alpha(m-m_0)}~ \nonumber \\
&& \times {\theta c^{\theta} \over (t+c)^{1+\theta}}~ 
{\nu L(m_i)^{\nu} \over (|\vec r - \vec{r_i}|+L(m_i))^{1+\nu}}
\Big] \, ,
\label{ETES}
\ea
where $L(m)=0.01\times 10^{0.5m}$ is the rupture length,
$p_m(m)$ is the magnitude distribution, $c$, $\theta$, $k$, $\alpha$, $\nu$ are 
adjustable parameters. Our method recovers the $\alpha$ value of the model
with an accuracy $\approx 0.03$, but underestimates the total number of 
aftershocks, because distant aftershocks outside the ``influence zone'' 
used for aftershock selection are missed (see results in Table 1).

\begin{figure}
\begin{center}
\includegraphics[width=0.5\textwidth]{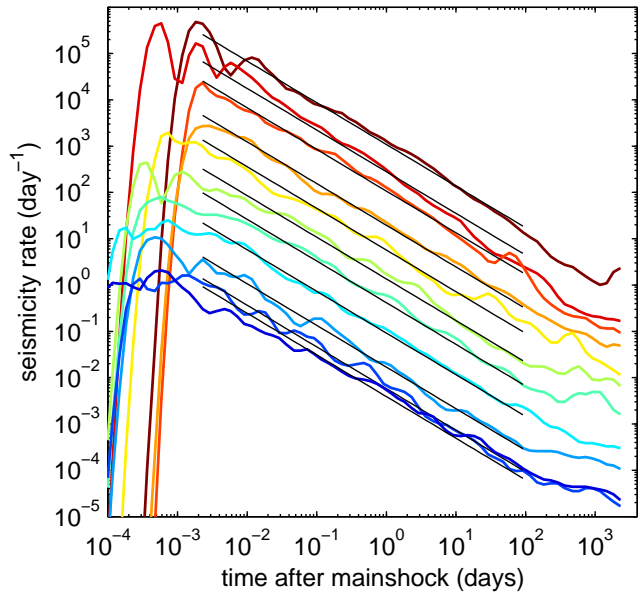}
\caption{\label{ANSS-1}
Same as Figure \ref{figa2} with different parameters 
$D_A=0.5$ km, $N_{L,A}=2$, $D_F=3$ km, $N_{L,F}=3$, and without including aftershocks 
in the influence zone of previous aftershocks. } 
\end{center}
\end{figure}

In order to study the rate of aftershocks at large times after a mainshock,  
we have selected aftershocks very close to the mainshock, in order to minimize 
the influence of non-correlated earthquake.
We used parameters $D_A=0.5$ km, $N_{L,F}=2$, $D_F=3$ km, 
and $N_{L,F}=3$, and we did not include aftershocks in the influence zone 
of previous aftershocks (see model \#8 in Table 1). We thus under-estimate 
the number of aftershocks for small mainshocks, because the influence zone 
of small events is smaller than the location accuracy, and because we miss 
more secondary distant aftershocks for small mainshocks.
The results are shown in Figure ~\ref{ANSS-1}. 
We observe that the duration of the aftershock sequence is at least 1000 days, 
independent of the mainshock magnitude, as predicted by the rate-and-state model 
of seismicity [{\it Dieterich}, 1994]. This figure also confirms that Omori's exponent 
does not depend on the mainshock magnitude. 

The instantaneous increase of seismicity rate after a mainshock, compared to the long-term rate, is at least of a factor $10^5$, maybe larger if we could detect aftershocks at shorter times. Using the rate-and-state model [{\it Dieterich}, 1994], this means that the ratio $\Delta \sigma /A \sigma_n$ 
(where $\Delta \sigma$ is the maximum Coulomb stress change, $A$ is a parameter of the 
rate-and-state friction law, and $\sigma_n$ is the normal stress) is at least 
equal to $\log(10^5)=11.5$. 
Assuming $A\sigma_n=0.4$ bar [{\it Toda and Stein}, 2003], this gives a maximum stress 
of $4.6$ bar.

\subsection{Magnitude distribution of triggered earthquakes \label{secpm}}

\begin{figure}
\begin{center}
\includegraphics[width=0.5\textwidth]{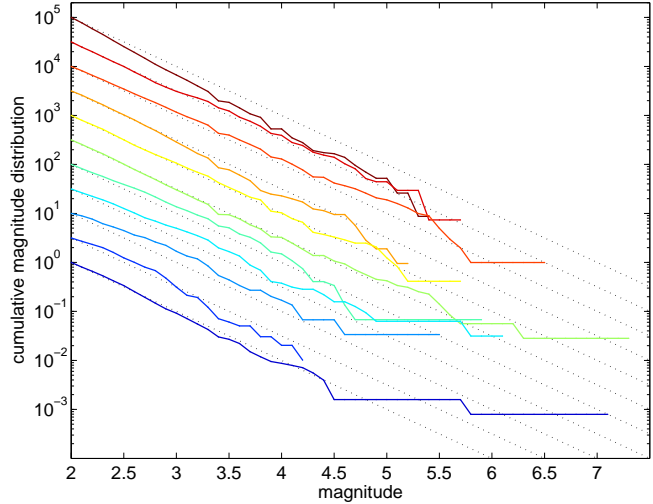}
\caption{\label{figPm} Cumulative magnitude distribution of triggered earthquakes,
for mainshock magnitudes ranging from 2 (below) to 7 (top) with a bin size of 0.5.
The curves have been normalized by a factor $10^{0.5m_M}$ for clarity. 
Dotted lines represent a GR law with $b=1$ for reference. 
%Duplicate events (which belong to several clusters within 
%the same bin of mainshock magnitude) have been removed. 
The number of events decreases with $m_M$ (between $N=11502$ for 
$7\leq m_M<7.5$ down to $N=1328$ for $7\leq m_M<2.5$) due to the smaller 
fraction of small earthquakes considered as mainshocks, and due to the 
shorter time window used for aftershock selection of small mainshocks.}
\end{center}
\end{figure}

We have analyzed the magnitude distribution of triggered earthquakes for each 
class of the mainshock magnitude between 2 and 7. As above, we select aftershocks 
in a time window [$t_{\rm min}(m_M)$, $t_{\rm max}(m_M)$] such that the catalog 
is complete above $m_d=2$ and that the influence of non-correlated earthquakes
is negligible ($\lambda(t,m_M)>0.1$ day$^{-1}$).
The results are shown in Figure \ref{figPm}. We observe that the 
magnitude distribution of triggered earthquakes follows the GR law with $b\approx1$, 
independent of the mainshock magnitude. For $7\leq m_M<7.5$ (Hector-Mine and Landers), 
the $b$-value seems larger than 1. However, if we use $m_d=3$ and a smaller minimum 
time $t_{\rm min}$, the magnitude distribution is close to the GR law with $b=1$.
These results imply that a small earthquake can trigger a much larger earthquake. 
It thus validates our hypothesis that the size of a triggered earthquake is not 
determined by the size of the trigger, but that any small earthquake can grow into 
a much larger one [{\it Kagan}, 1991b; {\it Helmstetter}, 2003; {\it Felzer et al.}, 2004].
The magnitude of the triggering earthquake controls only the number of triggered quakes.

\subsection{Proportion of triggered events in catalogs}

We use the scaling of the average number of triggered earthquakes per mainshock
and the hypothesis that the magnitude distribution of all quakes follows the GR 
law to derive the long-term fraction of aftershocks (assuming boundaries 
in time and space for the definition of aftershocks).
The rate of $m\geq m_d$ aftershocks triggered, directly or indirectly, 
by an earthquake of magnitude $m$ at time $t$ after this quake 
follows approximately
\be
\lambda_m(t)= { K_0~10^{\alpha m}~10^{- b m_d}~ \over t^p} \, ,
\label{lambdaM}
\ee
with $p=0.9$ and $K_0= 0.0041$ day$^{p-1}$. 
This relation holds at least for times $t>0.001$ day (86 sec), 
for all mainshock magnitudes, after correcting for the incompleteness
of the catalog at short times using eq. (\ref{mc}). 
For $m\geq 7$, the Omori law decay holds 
at least up to $T \approx 100$ days. If we assume that this ``aftershock duration'' 
does not depend on the mainshock magnitude, the average total number of $m>2$ 
aftershocks (including secondary aftershocks) triggered by $m\geq m_d$ mainshocks 
in the time window $c<t<T$ is
\be
N_{\rm aft}(m)=\int \limits_c^{T} \lambda_m(t)dt 
= K_0~10^{\alpha m - b m_d} {T^{1-p}-c^{1-p} \over 1-p} \, .
\label{NM}
\ee
Averaging over all magnitudes of the mainshock
we can compute the average total number $N_{\rm aft}$ of $m\geq m_d$ aftershocks 
(including secondary aftershocks) per mainshock of magnitude $m_d<m<m_{\rm max}$.
Assuming a GR law with $b=1$ and with an upper magnitude cut-off at $m=m_{\rm max}$,  $N_{\rm aft}$ 
is given by
\ba
N_{\rm aft} &=& \int\limits_{m_d}^{m_{\rm max}} p_(m)~ N_{\rm aft}(m)~ dm\\ \nonumber 
&=& K_0~{T^{1-p}-c^{1-p} \over 1-p}~\int\limits_{m_d}^{m_{\rm max}}
b~\ln(10)~10^{-b m}10^{\alpha m} dm \\ \nonumber 
&=& {K_0 b~10^{(\alpha-b) m_d}\over b-\alpha}
\left[1-10^{(\alpha-b)~(m_{\rm max}-m_d)}\right] 
~{T^{1-p}-c^{1-p} \over 1-p} \, .
\label{Naft}
\ea
The special case $\alpha=b$ gives
\be
N_{\rm aft} =K_0 b \ln(10)~(m_{\rm max}-m_d) ~ 
~{T^{1-p}-c^{1-p} \over 1-p} \, .
\label{Naft2}
\ee
% updated the results using line 2 in Table 1 
Using the parameters $\alpha=1.05$, $b=1.$, $c=0.001$ day, $K_0=0.0041$ day$^{p-1}$, 
and $m_d=2$ estimated for Southern California (see Table 1 line 2) and assuming a maximum 
magnitude $m_{\rm max}=8$ and an aftershock duration $T=100$ days, we obtain 
$N_{\rm aft}= 1.11$. In other words, a fraction $N_{\rm aft}/(N_{\rm aft}+1)=53$\% 
of $m\ge2$ earthquakes in California are aftershocks of a $m\ge 2$ earthquake. 
If we assume that Omori's law with $p=0.9$ holds up to $T=10000$ days (27 yrs), then 
$N_{\rm aft}= 2.06$ (67\% of earthquakes are aftershocks). 
%% replaced 1000 days by 10000 days (27 yrs)
The results do not 
depend on $c$ if $p<1$. The fraction of aftershocks increases if we include mainshocks 
with $m\le2$. Assuming that equation (\ref{NM}) and the GR law still hold down to 
$m_d=-1$ [{\it Abercrombie}, 1995], we obtain $N_{\rm aft}= 2$ for $T=100$ days: 
67\% of earthquakes of any 
magnitude are triggered by earthquakes of magnitude $m\geq -1$. The fraction 
of triggered events goes to 1 as the minimum magnitude $m_d$ goes to $-\infty$
[{\it Sornette and Werner}, 2004].

Our results are relatively close to the generic model of Californian 
aftershocks {\it Reasenberg and Jones}\ [1989, 1994] (RJ), which assumed
\be
\lambda_m(t)= { 10^{a+b(m-m_d)} \over (t+c)^p} \, ,
\label{lambdaMRJ}
\ee
with $c=0.05$ days, $p=1.08$, $b=0.91$, $a=-1.67$ [{\it Reasenberg and Jones}, 1994].
%with $c=0.05$ days, $p=1.07$, $b=0.9$, $a=-1.76$ (RJ89)
Their model gives $K_0=10^{a}=0.017$  day$^{-1}$ for $m_d=2$, but with 
smaller $\alpha$ and $b$ values ($\alpha$ is assumed to be equal to
$b$), so that the rate of $m\geq2$ aftershocks at $t=1$ day and for
$m=7$ is $\lambda_m(t)=758$ day$^{-1}$ for RJ model and $\lambda_m(t)=729$ 
% old version for RJ89:  $\lambda_m(t)=550$
day$^{-1}$ for our model (line \#2 in Table 1).

\section{Scaling of stress transfers with the mainshock magnitude}

Computing stress changes induced by a small earthquake at the location 
of another quake is very difficult. The stress field created by an earthquake 
is sensitive to the slip distribution on the fault, at least for small distances from 
the fault plane [{\it Steacy et al.}, 2004]. 
Slip distribution is usually available only for large $m\geq6$ earthquakes
in California. For small earthquakes, we can use a point-source model if we 
know the focal mechanism, to calculate stress for distances from the fault much 
larger than the rupture dimension. We can also estimate 
the most likely focal plane from the fault orientations in the area or 
from the orientation of the tectonic stress. Then we can model the rupture by 
a rectangular dislocation with a uniform or tapered slip. 
But this simple source model would still be incorrect close to the fault, where 
most aftershocks are located. 

{\it Marsan}\ [2004] solved this problem by computing the distribution of the 
stress induced by an earthquake at the location of another earthquake,
and by studying only the tail of the distribution (small absolute value of 
stress) corresponding to distances larger than the rupture length of the 
event. {\it Marsan}\ [2004] concluded that small earthquakes are at least as 
important as larger ones for redistributing stress between events.
This result is in agreement with our result for scaling the number 
of triggered earthquakes with the mainshock magnitude.
But even in the far field, these calculations of the Coulomb stress change 
induced by an earthquake are very inaccurate, because  the accuracy of focal 
mechanisms is on the order of 30$^\circ$ [{\it Kagan}, 2002b]. Another problem 
in [{\it Marsan}, 2004] is that he considers earthquakes of magnitude $m=3$, 
which have a rupture length $L\approx 300$ m, much smaller than the average 
accuracy on vertical coordinates of 4 km. These location errors can significantly 
bias any estimate of the stress change induced by small $m<5$ quakes 
[{\it Huc and Main}, 2003]. 

The observation that earthquake triggering scales with the magnitude
$m$ of the trigger as $\sim 10^{\alpha m}$ 
with $\alpha \approx b$ suggests that small 
events should not be neglected in studying stress interactions between
quakes.  However, directly calculating the stress induced by a small 
events is impossible due to the inaccuracy 
of earthquake locations and focal mechanisms. Even if we cannot compute the 
spatial distribution of the stress induced by a small earthquake, we can  
estimate how such small events trigger others. 
On average the total number of events triggered by an earthquake (integrated over 
space) will be independent of the earthquake source. 
For instance, we can account for the influence of small earthquakes in
the rate-and-state model of {\it Dieterich}\ [1994] by using for such small events a 
point source model which, as we know, depends only on the focal mechanism.
We can then compute the seismicity rate on each point of a grid by integrating 
on each cell that rate estimated using equation (12) of [{\it Dieterich}, 
1994] (using space-variable stress for a point-source model). 
If the size of each cell is equal to a few km, a point source model 
approximates $m\leq 4$ earthquakes well, because the integral of the seismicity 
rate over each cell depends only slightly on the geometry of the
rupture fault and the variations of fault slip.

Another alternative is to use empirical laws like the ETES model 
[{\it Kagan and Knopoff}, 1987b; {\it Felzer et al.}, 2003a; {\it Helmstetter et al.}, 
2004], rather than stress calculations and physical models of stress interactions.
We can estimate the spatial distribution of future aftershocks for a large 
earthquake by smoothing the locations of early aftershocks. A magnitude $m=7$ 
quake has enough aftershocks in the first hour to estimate the distribution of 
future aftershocks [{\it Helmstetter et al.}, 2004]. In contrast, predictions 
based on Coulomb stress change calculations require a good model of slip distribution 
on the fault, which is not available until a few hours after the mainshock at best.

\section{Spatial distribution of seismicity \label{secD}}

\begin{figure}
\begin{center}
\includegraphics[width=1\textwidth]{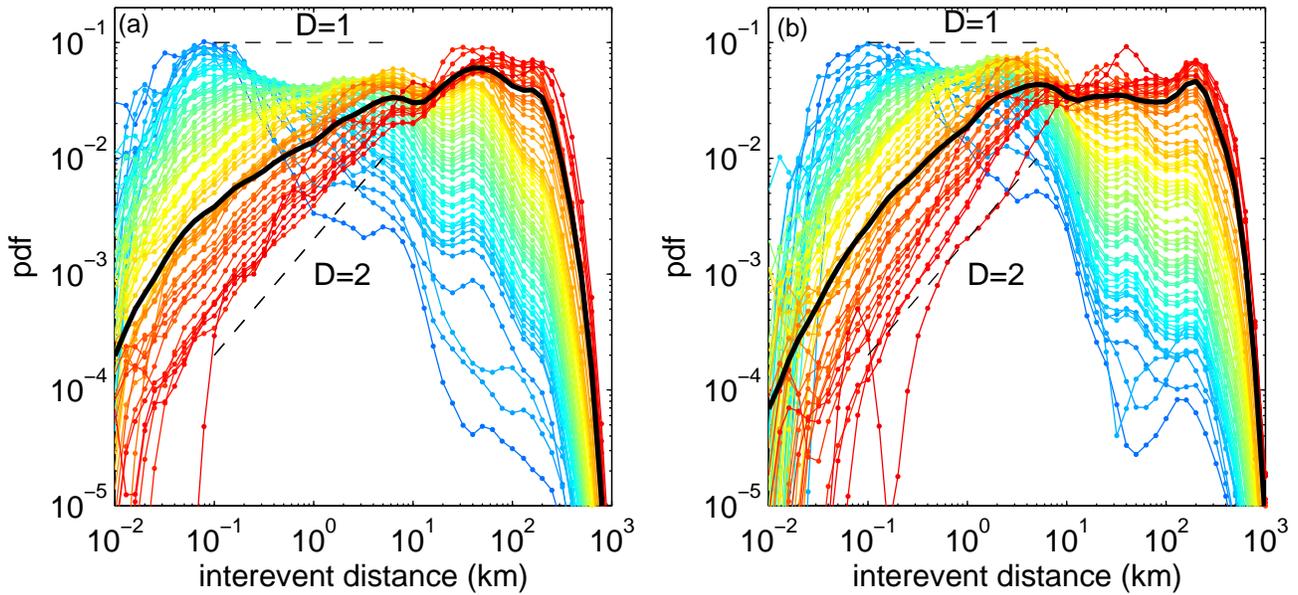}
\caption{\label{pdfrtS}
Distribution of distances between hypocenters $p_r(r,t)$ for SHLK (a) and 
HCS (b) catalogs,
using only earthquake pairs with inter-event times in the range [$t$, $1.25t$],
where $t$ increases between 1.4 minutes (blue curve) to 2500 days (red curve).
The fractal dimension of $p_r(r)$, measured for $0.1\leq r \leq 5$ km
for all earthquakes (black lines), is $D=1.54$ for SHLK catalog and $D=1.73$
for HCS catalog. }
\end{center}
\end{figure}

We have estimated the distribution of distances between hypocenters
$p_r(r)$, using the relocated catalogs for Southern California by
{\it Hauksson et al.}\ [2003] (HCS) and {\it Shearer et al.}\ [2003] (SHLK). 
These catalogs apply waveform cross-correlation to obtain precise differential 
times between nearby events. These times can then be used to greatly 
improve the relative location accuracy within clusters of similar events. 
Locations in these two catalogs do not always agree in detail, reflecting
their different modeling assumptions and seismic velocity structures. However, 
their overall agreement is quite good, particularly when compared to the 
standard catalog locations. In many regions, these new catalogs resolve 
individual faults in what previously appeared to be diffuse earthquake clouds. 
In both catalogs, we have selected only $m\geq2$ earthquakes
relocated with an accuracy of $\epsilon_h$ and $\epsilon_z$ 
smaller than 100m. In the HCS catalog, there are 71943 $m\geq2$ earthquakes
between 1984 and 2002, out of which 24127 (34\%) are relocated 
with $\epsilon_h<0.1$ km and $\epsilon_z<0.1$ km. In the SHLK catalog, there 
are 82442 $m\geq2$ earthquakes in the same period [1984, 2002], out of which
33676 (41\%) are relocated with $\epsilon_h<0.1$ km and $\epsilon_z<0.1$ km.
The probability density function of distances $p_r(r)$ between hypocenters is close 
to a power-law $p_r(r \sim r^{D-1}$ in the range $0.1\leq r \leq5$ km.
The correlation fractal dimension (measured by least-square linear regression 
of $\log(r)$ and $\log(p_r(r))$ for $0.1\leq r \leq5$ km)
 is $D=1.54$ for SHLK and $D=1.73$ for HCS 
(see black lines in Figure \ref{pdfrtS}).
The faster decay for $r<0.1$ km is due to location errors, and the roll-off for 
distances $r>5$ km is due to the finite thickness of the seismogenic crust. 
The difference in the $D$-value between the two catalogs may come from larger 
location errors in HCS, but also from errors in estimating the $D$-value 
(see {\it Kagan}\ [2004c] for a discussion of errors and biases in determining 
the fractal correlation dimension). 

To check the accuracy of the $D$-value, 
we tested a synthetic catalog, generated with the same 
number of events as the real catalog. We used random longitudes and latitudes, 
with a uniform epicentral distribution within the same boundaries as real data. 
Depths in the synthetic catalog were chosen by shuffling the 
depths of earthquakes from the real catalog to keep the same 
depth distribution. The fractal dimension in the range $0.1\leq r\leq5$ km was 
$D=2.93 \pm 0.04$: a little smaller than the value $D=3$ expected for a purely 
uniform distribution in a volume. This discrepancy is due to three factors: the finite number 
of events, the non-uniform distribution of depths, and  finite size effects. 

The spatial clustering of earthquake hypocenters measured for the full catalog
is due both to the fault network structure  and to earthquake interactions.
The latter modify long-term spatial distribution by increasing
the fraction of small inter-event distances, that is decreasing the fractal 
dimension [{\it Kagan}, 1991a]. The clustering of aftershocks around  mainshocks, 
and of secondary aftershocks close to the direct aftershock of the mainshocks, and 
so on, can create a fractal distribution. This results from the scaling of the aftershock 
zone with magnitude coupled with the GR law, without any underlying fractal fault 
structure [{\it Helmstetter and Sornette}, 2002]. 

Over very long time intervals, 
triggered earthquakes should be located uniformly on the fault network,
and the fractal dimension of the whole catalog should reach a constant value.
In our model described in equation (\ref{nl}), $p_r(r)$ represents the
``long-term'' time-independent spatial distribution of seismicity.
Distribution $p_r(r)$ gives the average density of earthquakes at a distance
$r$ from the mainshock due to fault structure geometry and 
the fault's time-independent heterogeneity. This does not take into 
account earthquake interactions. The effect of the mainshock, which increases 
or decreases the seismicity rate at close distances by modifying the stress,
is described in the factor $f(\sigma,t)$ in (\ref{nl}). 
Therefore, we should not use in (\ref{nl}) the function $p_r(r)$ estimated 
from the catalogs of Southern California seismicity because these 
catalogs cover less than 20 years and contain a majority of aftershocks. 
If we had included that function, we would be double-counting earthquake interactions.

\begin{figure}
\begin{center}
\includegraphics[width=0.5\textwidth]{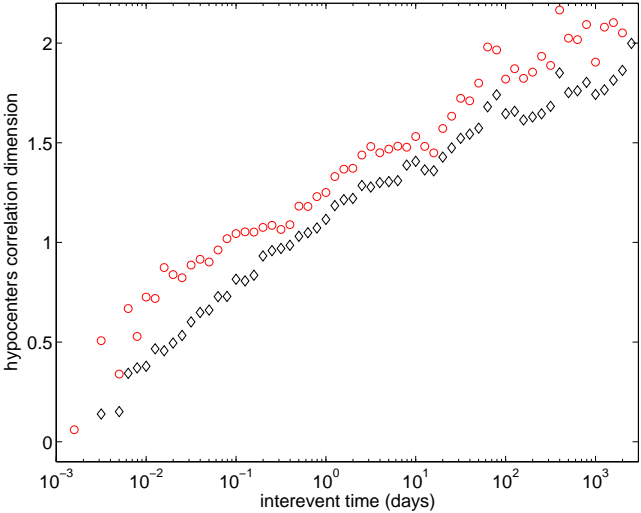}
\caption{\label{pdfrtSDt}
Fractal dimension of $p_r(r,t)$ as a function of inter-event time $t$, using SHLK catalog (diamonds) and HCS catalog (circles).}
\end{center}
\end{figure}

To estimate and remove the time dependence of the spatial 
distribution of inter-event distances, we have measured that distribution  
$p_r(r,t)$ using only earthquake pairs with an inter-event 
time $\tau$ in the range [$t$, $t+dt$]. 
%We have also measured the distribution 
%$p_r(r,\tau\geq t) =\int_t^\infty p_r(r,t')dt'$ cumulated over 
%all times larger than $t$ (i.e., using only events with inter-event times larger 
%than $t$). 
The results are shown in Figure \ref{pdfrtS}. As the minimum inter-event time 
increases, the fraction of small distances will decrease. The fractal dimension of 
%the cumulative distribution $p_r(r,\tau\geq t)$ increases with $t$ from the value 
%measured for the whole catalog to a maximum value close to 2 for inter-event 
%times larger than 1000 days. For 
$p_r(r,t)$ increases between $D\approx0$ at times $t=5$ minutes up to $D\approx2$ 
for $t=2500$ days (see Figure \ref{pdfrtSDt}). 
 This maximum inter-event time of 2500 days is long enough so 
that earthquake interactions are negligible compared to tectonic loading, and
a very small fraction of earthquake pairs with $\tau>1000$ days belong to the same aftershock sequence. 
This value $D=2$, measured for $t=2500$ days, can thus be interpreted as the fractal
dimension of the active fault network.

\section{Conclusion}

Although large earthquakes are much more important than smaller ones for 
energy release, small quakes have collectively the same influence as large ones %%
for stress changes between earthquakes, due to seismic spatial clustering.
Because the stress drop is constant, the stress change induced by 
an earthquake of magnitude $m$ at the location of other earthquakes 
(located on a fractal network of dimension $D$) increases
with $m$ as $\sim10^{Dm/2}$. Measuring directly the scaling of 
such stress change with magnitude is difficult, because the accuracy of 
earthquake locations and focal mechanisms is limited.
We can, however, estimate indirectly the relative importance of small and large 
events for stress transfer by studying the properties of
triggered seismicity. If quakes are triggered by static stress
changes, then the number of such triggered events should also scale as 
$\sim  10^{Dm/2}$ with the magnitude of the trigger.

We have measured the average seismic rate triggered by an earthquake
for Southern California seismicity. We found that the rate of triggered
events decays with time according to Omori's law $\sim 1/(t+c)^p$ 
with $p=0.9$ and $c<3$ minutes (after correcting for the increase in the 
completeness threshold after a large mainshock, see eq.~\ref{mc}). This decay 
is independent of the mainshock magnitude $m$ for $2<m<7.5$. We also found that 
the magnitude of triggered aftershocks follows the Gutenberg-Richter law
with $b=1$ and is independent of the mainshock magnitude.
The rate of triggered quakes increases with $m$ as $\sim
10^{\alpha m}$, with an exponent $\alpha=1.05 \pm 0.05$.

We have measured the fractal correlation dimension $D$ for two
different catalogs of relocated earthquakes in Southern California,
given by the exponent of the cumulative distribution of the distances
between all pairs of hypocenters. This fractal dimension $D\approx 1.6$, 
measured in a limited time-window, characterizes the spatial clustering of 
earthquakes due to fault network structure and earthquake 
interactions. By using only earthquake pairs with inter-event times larger 
than a threshold to compute $D$, we have tried separating the effects that fault
geometry and earthquake triggering have upon spatial clustering. 
For inter-event times larger than 1000 days, we obtain $D\approx2$.
Thus, our result $\alpha \approx D/2$ supports the assumption that
earthquakes are triggered by static stress from past events. 

The fact that $\alpha$ is nearly equal to the $b$-value has important 
consequences for earthquake triggering. It means that equal magnitude bands 
contribute equally to stress at the location of a future hypocenter, e.g.,
earthquakes with $2<m<5$ have collectively the same importance as $5<m<8$ 
events  for stress changes and for triggering, because the frequency of small 
earthquakes compensates for their lower triggering potential. 
Even if explicit stress calculations based on cataloged earthquakes
have limited accuracy, we can estimate how small earthquakes affect 
earthquake triggering. This can be done with a spatial resolution of
a few km, using an ETES model or the more physical rate-and-state model.

\begin{acknowledgments}

We acknowledge the Advanced National Seismic System, Egill
Hauksson, Peter Shearer, and the Harvard group for the
earthquake catalogs used in this study. 
We thank Emily Brodsky, Jim Dieterich, Karen Felzer, Paolo Gasperini, Jean-Robert
Grasso, and Max Werner for useful discussions. 
We are grateful to the associate editor Massimo Cocco, and to 
the reviewers Bruce Shaw, Susanna Gross and Ruth Harris for
 useful suggestions.
We acknowledge David Marsan for sending us a preprint 
of his paper. We are grateful to Kathleen Jackson for editing the manuscript.
This work is partially supported by NSF EAR 0409890 and by the
Southern California Earthquake Center (SCEC). 
SCEC is funded by NSF Cooperative Agreement EAR-0106924 and
USGS Cooperative Agreement 02HQAG0008.
The SCEC contribution number for this paper is 805.
\end{acknowledgments}

\end{article}

\newpage

\begin{table}
\label{tab1}
\caption{Scaling of aftershock productivity and earthquake 
frequency with magnitude.
%, for Southern California seismicity and for 
%synthetic ETES catalogs (see eq. (\ref{ETES}), generated with parameters $b=1$, 
%$\theta=0.1$, $m_0=0.5$, $m_{\rm max}=8$, $c=0.001$ day, $\nu=3$ and
%$\epsilon=1.5$ km (added uniform noise on locations). 
Times are in days and distances in km.
Parameters $\Delta m$, $D_F$, $D_A$, $N_{L,F}$, $N_{L,A}$ and $T_F$ 
are used in the selection of mainshocks and aftershocks (see section \ref{secm}).
Parameters $t_{\rm min}$, $t_{\rm max}$ and $\lambda_{\rm min}$ define
the time interval used to estimate the scaling of the number of
aftershocks with the mainshock magnitude (see section \ref{seca}).
$N$ is the total number of events with $m\geq m_d$ and $N_{ms}$ is the number 
of mainshocks. $K_0$ defined by (\ref{Kmm}) measures aftershock productivity.} 
\begin{flushleft}
\begin{tabular}{lrcrrrrrrlclrrrl}
\# & $m_d$ & time & $\Delta m$ & $D_F$ & $D_A$ & $N_{L,F}$ & $N_{L,A}$ & $T_F$ &
$t_{\rm min}$ & $t_{\rm max}$ & $\lambda_{\rm min}$ & $\alpha$ & $N$ & $N_{ms}$ & $K_0$ \\
\tableline\\
%\multicolumn{16}{c}{California} \\
1      & 2 & 1980-2004 & 1 & 3 & 2   & 3 & 2 & 365 & 0.0020 &  10 & 0.100 & 0.99 & 101402 & 19247 & 0.0092\\
2$^c$  & 2 & 1980-2004 & 1 & 3 & 2   & 3 & 2 & 365 & 0.0020 &  10 & 0.100 & 1.05 & 101402 & 19247 & 0.0041\\
3$^c$  & 2 & 1980-2004 & 1 & 3 & 2   & 3 & 2 &  36 & 0.0020 &  10 & 0.100 & 1.02 & 101402 & 33123 & 0.0065\\
4$^c$  & 2 & 1980-2004 & 1 & 3 & 2   & 3 & 2 & 730 & 0.0020 &  10 & 0.100 & 1.06 & 945736 & 15130 & 0.0035\\
5$^c$  & 2 & 1980-2004 & 0 & 3 & 2   & 3 & 2 & 365 & 0.0020 &  10 & 0.100 & 0.95 & 101402 & 26004 & 0.0167\\
6$^c$  & 2 & 1980-2004 & 2 & 3 & 2   & 3 & 2 & 365 & 0.0020 &  10 & 0.100 & 1.09 & 101402 & 18784 & 0.0024\\
7$^c$ & 2 & 1980-2004  & 1 & 3 & 3   & 3 & 3 & 365 & 0.0020 &  10 & 0.500 & 1.04 & 101402 & 19247 & 0.0048\\
8$^{a,c}$ & 2 & 1980-2004 & 1 & 3 & 0.5 & 3 & 2 & 365 & 0.0020 &  10 & 0.005 & 1.16 & 101402 & 19247 & 0.0009\\
9$^{a,c}$& 2 & 1980-2004 & 1 & 3 & 0.5 & 3 & 2 & 365 & 0.0020 & 100 & 0.000 & 1.15 & 101402 & 19247 & 0.001\\ 
10$^{a,c}$& 2 & 1980-2004 & 1 & 3 & 0.5 & 3 & 2 & 365 & 0.0020 &1000 & 0.000 & 1.12 & 101402 & 19247 & 0.0013\\
%A1a$^c$d&2 & 1984-2004 & 1 & 2 & 0.2 & 3 & 2 & 365 & 0.0020 & 1000& 0.000 & 0.95 &  88618 &  1818 & 0.187\\ %ANSS er<0.1km
11$^c$ & 2 & 1980-2004 & 1 & 1 & 4   & 1 & 4 & 365 & 0.0020 &  10 & 5.000 & 0.94 & 101402 & 37839 & 0.0193\\
12$^c$ & 2 & 1980-2004 & 1 & 1 & 2   & 1 & 2 & 365 & 0.0020 &  10 & 2.000 & 1.01 &  98805 & 36864 & 0.0078\\ 
13$^c$ & 2 & 1980-2004 & 1 & 4 & 1   & 4 & 1 & 365 & 0.0020 &  10 & 0.100 & 1.13 & 101402 & 15153 & 0.001 \\
14$^c$ & 2 & 1980-2004 & 1 & 4 & 4   & 4 & 4 & 365 & 0.0020 &  10 & 0.100 & 1.02 & 101402 & 15153 & 0.0064\\
15$^c$ & 2 & 1980-2004 & 1 & 1 & 1   & 1 & 1 & 365 & 0.0020 &  10 & 0.500 & 1.05 & 101402 & 37839 & 0.003 \\ 
16$^c$ & 2 &1980-1992.4& 1 & 3 & 2   & 3 & 2 & 365 & 0.0020 &  10 & 0.100 & 1.00 &  40565 &  9603 & 0.0056\\ 
17$^c$ & 2 &1992.4-2004& 1 & 3 & 2   & 3 & 2 & 365 & 0.0020 &  10 & 0.100 & 1.05 &  60837 &  9187 & 0.0041\\ 
18     & 3 & 1980-2004 & 1 & 3 & 2   & 3 & 2 & 365 & 0.0003 &  10 & 0.010 & 1.07 & 101402 & 19247 & 0.0033\\ %h=0.2 
19$^c$ & 3 & 1980-2004 & 1 & 3 & 2   & 3 & 2 & 365 & 0.0020 &  10 & 0.010 & 1.10 & 101402 & 19247 & 0.0023\\ %h=0.1
20$^{b,c}$&2&1980-2004 & 1 & 3 & 3   & 4 & 2 &  36 & 0.0020 &  10 & 0.500 & 1.01 & 101402 & 29210 & 0.0075\\ 
21$^c$& 2 &  1980-2004 & 1 & 3 & 3   & 4 & 2 &  36 & 0.0020 &  10 & 0.500 & 1.02 & 101402 & 32801 & 0.006 \\ 
% ETES synthetic catalogs
%\multicolumn{16}{c}{ETES, $\alpha=0.5$, $k=0.018$}\\ %ETES.5-1, M0=0.5
%23$^c$  & 2 & 1980-2004 & 1 & 4 & 2 & 4 & 2 & 36 &0.002 & 10. & 0.5 & 0.52 & 125350 & 83379 & 0.15\\
%\multicolumn{16}{c}{ETES, $\alpha=0.9$, $k=0.0042$}\\ %ETES.3-1
%24$^c$  & 2 & 1980-2004 & 1 & 4 & 2 & 4 & 2 & 36 &0.002 & 10. & 0.5 & 0.91 & 141866 & 62817 & 0.026\\
%\multicolumn{16}{c}{ETES, $\alpha=1$, $k=0.0022$}\\ %ETES.6-2, M0=0.5
%25 & 2 & 1980-2004 & 1 & 4 & 2 & 4 & 2 & 36 &0.002 & 10. & 0.5 & 1.02 & 161351 & 45589 & 0.012\\
\tableline
\end{tabular}
\end{flushleft}
\end{table}

%\end{article}

\newpage

\newpage

\newpage

\newpage

\end{document}